\documentclass[aps,prl,twocolumn,preprintnumbers,groupedaddress,nofootinbib]{revtex4}

\usepackage[latin1]{inputenc}
\usepackage[english]{babel}
\usepackage{amsfonts}
\usepackage{amsmath}    
\usepackage{amssymb}
\usepackage{amsthm}
\usepackage[pdftex]{color}
\usepackage{slashed}
\usepackage{graphicx}

\usepackage[pdftex,colorlinks,pdfpagelabels]{hyperref}
\usepackage[figure,table]{hypcap} 
\hypersetup{
   bookmarksnumbered,
   pdfstartview={FitV},
   pdfpagemode={UseOutlines},
   pdfauthor={Jasper Hasenkamp},
   pdftitle={Neutrino Self-Interactions},
   pdfsubject={scientific publication}
   ,citecolor={blue},
   linkcolor={blue},
   urlcolor={blue},
   filecolor={blue}
} 

\def\a{\alpha}

\def\d{\delta}
\def\e{\epsilon}

\def\g{\gamma}

\def\m{\mu}
\def\n{\nu}

\def\s{\sigma}

\def\D{\Delta}

\def\L{\Lambda}
\def\O{\Omega}

\newcommand{\MeV}{\text{ MeV}}
\newcommand{\GeV}{\text{ GeV}}
\newcommand{\TeV}{\text{ TeV}}

\newcommand{\mplanck}{\ensuremath{M_{\text{pl}}}}

\newcommand{\Uoney}{\ensuremath{U(1)_Y}}
\newcommand{\Uonex}{\ensuremath{U(1)_X}}

\newcommand{\fb}[3]{\left(\frac{#1}{#2}\right)^{#3}}
\newcommand{\order}[1]{\ensuremath{\mathcal{O}\left(#1\right)}}

\newcommand{\be}{\begin{equation}}
\newcommand{\ee}{\end{equation}}

\begin{document}

\date{\today}

\title{ \bf
Neutrino Self-Interactions 
}
%
\author{Jasper Hasenkamp}
\email{Jasper.Hasenkamp@gmail.com}
\affiliation{Hamburg, Germany.}

\begin{abstract}
\noindent
We propose a theory that equips the active neutrinos with interactions among themselves that are at least three orders of magnitude stronger than the weak interaction.
We introduce an Abelian gauge group $\Uonex$ with vacuum expectation value $v_x \lesssim \order{100\MeV}$.
An asymmetric mass matrix implements the active neutrinos as massless mass eigenstates carrying "effective" charges.
To stabilize $v_x$, supersymmetry breaking is mediated via loops to the additional sector with the only exception of xHiggs terms. No Standard Model interaction eigenstate carries $\Uonex$ charge. Thus the dark photon's kinetic mixing is two-loop suppressed.
With only simple and generic values of dimensionless parameters, our theory might explain the high-energy neutrino spectrum observed by IceCube including the PeV neutrinos. We comment on the imposing opportunity to incorporate a self-interacting dark matter candidate.
\end{abstract}

\maketitle

\thispagestyle{empty}
\vspace{1cm}


\section{Introduction}
The observation of high energy neutrinos in the IceCube detector~\cite{Aartsen:2013bka,Aartsen:2013jdh} marked the beginning of extragalactic high-energy neutrino astronomy. After three years of data-taking, 37 events with energies between 30 TeV and 2 PeV provide evidence at 5.7$\sigma$ for the existence of an extraterrestrial neutrino flux~\cite{Aartsen:2014gkd}.

This discovery will tell us much more about the mysterious nature of neutrinos than the detection of neutrinos from Supernova 1987A, because of the longer distance, higher energy and higher number of events. 
The observed neutrino energy spectrum is 
well described by a simple power law with spectral index $-2.49 \pm 0.08$~\cite{Aartsen:2015zva}.
Some proposed astrophysical models might be consistent with these observations~\cite{Loeb:2006tw}, while certainly none is compelling~\cite{Aartsen:2014muf}.

The most interesting aspect of the spectrum is the lack of any event at energies above 2 PeV. Today, there is no significant statistical preference (1.2 $\s$) for a cut-off. However, its presence can likely be determined with additional data in the near future~\cite{Aartsen:2015zva}. At the same time the observed neutrino flux is comparable to the Waxmann-Bahcall bound~\cite{Waxman:1998yy}, i.e., the cosmogenic neutrino flux at EeV energies produced by ultra high-energy cosmic-ray protons. Since different processes are expected at PeV and EeV energies, it has been identified as a "coincidence problem" if they shall give almost the same flux~\cite{Ioka:2014kca}.

It has been proposed that this coincidence (and the cut-off at 2 PeV) could be due to a neutrino self-interaction $\nu$SI, that implies EeV cosmogenic neutrinos to loose energy on their way to Earth in scatterings with cosmic background neutrinos~\cite{Ng:2014pca,Ioka:2014kca}.
This "cascade mechanism" requires an interaction strength $g_\nu /m_x \sim 5 \GeV^{-1}$ with effective coupling constant $g_\nu$ for mediating boson masses $10 \MeV \lesssim m_x \lesssim 100 \MeV$. This is at least three orders of magnitude stronger than the weak interaction.
Interestingly enough, at present such strong $\nu$SI are not in conflict with any observation.
In addition, the $\nu$SI can explain the lack of events with energies between 400 TeV and 1 PeV or any possible gap feature, that might become evident with future data.

However, neutrinos form a $SU(2)$ doublet with the charged leptons, while especially the electron is subject to very strong constraints on any "secret" interaction~\cite{Chiang:2012ww}. %
As a consequence, neutrinos can receive a sizeable effective charge via mass mixing only, while the order of their sub-eV masses is tiny compared to any other mass in the Standard Model of particle physics (SM), see for example~\cite{Babu:2003is,Pospelov:2011ha,Heeck:2012bz,Cherry:2014xra}.
Furthermore, every particle physics model with a light mediator faces the same inherent problem: the instability of its scale to quantum corrections, which is closely connected to the infamous hierarchy problem.
Nevertheless, due to the fascinating possibilities of neutrino physics to resolve at first sight unconnected puzzles~\cite{Aarssen:2012fx,Bringmann:2013vra}, efforts are increasing recently~\cite{Dasgupta:2013zpn,Ko:2014bka,Archidiacono:2014nda,Cherry:2014xra,Kouvaris:2014uoa}.

In this work, we present for the first time a consistent theory of $\nu$SI.  
We propose a supersymmetric theory amending the SM by an Abelian gauge group $\Uonex$, that is spontaneously broken.
We show as a proof of principle, that the cascade mechanism can be implemented successfully, even though, theoretical and observational requirements are partly in fundamental tension.

Key to our work are two notions: Firstly, an asymmetric neutrino mass matrix. In this way, the active neutrinos receive an "effective" charge via mass mixing, while they can stay massless. We believe that this will become the standard mechanism to introduce $\nu$SI. 
Secondly, with the exception of additional soft "xHiggs" terms, supersymmetry (SUSY) breaking is mediated to the additional "xsector" via loop corrections only. These are controlled by small Yukawa couplings, so that radiative stability is ensured. Our theory is UV complete.

Thanks to the minimality of our theory, we can point out generic features and implications. For example, mixing of the additional vector boson with the photon arises at two-loop order. So it might be called "very dark", c.p.~\cite{Fradette:2014sza}. We comment on the imposing opportunity that the $\nu$SI is shared by the cold dark matter and/or hot dark matter.

The letter is organised as follows: We introduce our theory in Sec.~\ref{sec:theory}, where we also discuss its radiative stability and the sparticle mass spectrum. In Sec.~\ref{sec:eff}, we provide the mechanism for effective neutrino charges and demonstrate how the cascade mechanism can be implemented. The resulting thermal history of the Universe is presented in Sec.~\ref{sec:thermal}, where we also discuss the phenomenology of the lightest supersymmetric particle. We summarise and conclude in the final section.

\section{Theory and setup}
\label{sec:theory}
We consider the extension of the SM gauge group, 
$G_\text{SM} = SU(3)_c \times SU(2)_{L} \times \Uoney$, by an Abelian gauge 
symmetry \Uonex\ with corresponding gauge boson $X$ and gauge coupling $g_x$.
Furthermore, we assume that supersymmetry (SUSY) is a spontaneously broken symmetry in Nature.
Note that taken for itself the one and only motivation to assume SUSY in this work is the stabilisation of VEVs against quantum corrections. If SUSY is not realised in Nature or in a very different way than thought today and the (general) hierarchy problem is absent, we could drop that assumption without any loss of motivation or validity.

We introduce a Dirac ``x-neutrino'' $\nu_x = (\nu_L^x \text{, } \nu_R^x)^T$, which is neutral under $G_\text{SM}$ but carries 
\Uonex\ charge, while the particles of the Minimal Supersymmetric Standard Model (MSSM) are neutral under \Uonex. We further add a Dirac sterile neutrino $\nu_s = (\nu_L^s \text{, } \nu_R^s)^T$, which is neutral under all gauge interactions. 
The additional chiral supermultiplets with their corresponding charges are listed in Tab.~\ref{tab:chiral} and the additional gauge supermultiplet in Tab.~\ref{tab:gauge}.

The superpotential of our theory reads
\begin{align}
\label{W}
W=& \, W_\text{mssm} + y_L \bar\nu_R^s L H_u - y_{+}\, \bar\nu_R^s \nu_L^x \phi_x^+ - y_{-} \, \bar\nu_R^x \nu_L^s  \phi_x^- \nonumber \\
&+ \frac{\m_s}{2}  \bar\nu_R^s \nu_L^s + \frac{\m_x}{2}  \bar\nu_R^x \nu_L^x 
- \m_\phi \phi_x^+ \phi_x^-  \, ,
\end{align}
where $W_\text{mssm}$ denotes the MSSM superpotential. Additional terms are the standard right-handed neutrino term, two Yukawa terms corresponding to the two additional xHiggs fields, mass terms for the sterile neutrino and the xneutrino as well as a $\m$-term for the xHiggses.
Holomorphicity requires two different xHiggses, $\phi_x^+$ and $\phi_x^-$, to build the two corresponding Yukawa terms.
Note that non-supersymmetric models could contain only one xHiggs field.
Gauge-invariance forbids any Higgs-xHiggs mixing terms, also known as Higgs portal. 
Note that Majorana mass terms for $\nu_R$ and $\nu_L$ are allowed, since both are singlets under all gauge groups. The addition of such masses represents the straightforward way to provide SM neutrinos with masses. As this is not in the focus of this paper and would work out straightforwardly, we omit them for simplicity. As technical justification one could assume a conserved lepton number or simply that the terms are negligible.
Electroweak symmetry breaking occurs unaffected. At low energy the 
 vacuum expectation value (VEV) $v_\text{ew}$ of the SM Higgs $h$ breaks ${G}_\text{SM}$ as usual.
 The theory is anomaly free, since Dirac neutrinos do not introduce anomalies.
R-charges are unambiguous and reduce  the number of possible superpotential terms favourably.  
 
%
%

\begin{table}
\centering
 \begin{tabular}{c |c| c| c}
 Name(s) & Scalar  &  Fermion & X charge  \\ \hline
rh.\@ sterile (s)neutrino $\bar\nu_R^s$ & $(\widetilde{\nu}_R^s)^\ast$  & $(\nu_R^s)^\dagger$  &  0 \\
lh.\@ sterile (s)neutrino $\nu_L^s$ & $\widetilde{\nu}_L^s$  & $\nu_L^s$  &  0 \\
rh.\@ x(s)neutrino $\bar\nu_R^x$ & $(\widetilde{\nu}_R^x)^\ast$  & $(\nu_R^x)^\dagger$  &  $q_x$ \\
lh.\@ x(s)neutrino $\nu_L^x$ & $\widetilde{\nu}_L^x$  & $\nu_L^x$  &  $-q_x$ \\
+xHiggs(ino) $ \phi_x^+$ &  $ \phi_x^+$ & $  \widetilde{\phi}_x^+ $ & $q_x$ \\
-xHiggs(ino) $ \phi_x^-$ &  $ \phi_x^-$ & $  \widetilde{\phi}_x^- $ & $-q_x$
\end{tabular}
\caption{Additional chiral supermultiplets. All additional multiplets are SM singlets. Superpartners are denoted by a tilde. Their names are obtained by addition of an ``s'' or ``ino'' as given in the first column. There are complex scalars and left-handed, two-component Weyl fermions. Suggestively, the neutrino names comprise either right-handed (rh.) or left-handed (lh.). 
}
\label{tab:chiral}
\end{table}
\begin{table}
\centering
 \begin{tabular}{l|c |c| c }
 Name(s) & Vector  &  Fermion & X charge  \\ \hline
X(ino) & $X$ & $\widetilde{X}$  & 0  
\end{tabular}
\caption{The additional, SM-singlet gauge supermultiplet. The Xino is denoted by a tilde.
}
\label{tab:gauge}
\end{table}
%
%
%

Spontaneous $\Uonex$ breaking requires a negative mass-squared for an xHiggs. Therefore, we consider, in addition to the MSSM soft SUSY breaking terms $\mathcal{L}^\text{mssm}_\text{soft}$, the following soft SUSY breaking xHiggs terms
\begin{align}
\mathcal{L}_\text{soft}&=\mathcal{L}_\text{soft}^\text{mssm} - m^2_{\phi_x^+} (\phi_x^+)^\ast \phi_x^+ - m^2_{\phi_x^-} (\phi_x^-)^\ast \phi_x^+ \nonumber \\ 
&- (b_\phi \phi_x^+ \phi_x^- + \text{c.c.}) .
\end{align}

%
In analogy to well-studied cases, we note shortly that
for $|\m_\phi|^2 + m^2_{\phi_x^+} < 0$ and $|\m_\phi|^2 + m^2_{\phi_x^-} < 0$, respectively, the xHiggses will acquire VEVs $\langle \phi_x^+ \rangle $ and $\langle \phi_x^- \rangle$. One degree of freedom becomes a Goldstone boson and
 the X vector boson mass is determined as $m_X^2=  q_x^2 g_x^2 (\langle \phi_x^+ \rangle^2 + \langle \phi_x^- \rangle^2)= q_x^2 g_x^2 v_x^2$, where we introduced the shorthand notation $v_x^2 = \langle \phi_x^+ \rangle^2 + \langle \phi_x^- \rangle^2 $.
%
For the time being, we assume spontaneous symmetry breaking with all dimensionful parameters of the same order of magnitude, i.e., $|\m_\phi|^2 \sim  m^2_{\phi_x^+}\sim m^2_{\phi_x^-} \sim b_\phi$ and so $  \langle \phi_x^+ \rangle \sim \langle \phi_x^- \rangle \sim |\m_\phi|^2$, so that also the masses of the xHiggses and xHiggsinos will be of the order of $m_X$. 
%

It is interesting to note that non-supersymmetric theories allow to set an arbitrarily small tree-level xHiggs mass by choice of the size of the corresponding parameter in the Lagrangian. While SUSY takes away that freedom, we will see below that xHiggs masses seem to be phenomenologically bounded from below to values at least of the order of $m_X$ anyway. In this sense, SUSY provides a favourable, natural reason for the size of the xHiggs mass(es).

\textbf{Radiative stability}
Radiative stability of the (small) VEVs requires that quantum corrections to the xHiggs masses $\d m_{\phi_x^\pm}^2$ are smaller than their tree-level masses, $\d m_{\phi_x^\pm}^2 < m_{\phi_x^\pm}^2$.\footnote{
To reduce clutter, we use the same symbol $\phi_x^\pm$, here.
}
%
%
The leading (one-loop) correction is due to $\bar{\n}^s_R$-$\n_L^x$ and $\bar{\n}^s_L$-$\n_R^x$ in the loop that couple via the third and forth term in~\eqref{W} to $\phi_x^+$ and $\phi_x^-$, respectively. It is 
\be
|\d m_{\phi_x^\pm}| \sim y_\pm \D m_{\tilde \n_R^s}/(\sqrt{8} \pi).
\ee
If SUSY were unbroken, quantum corrections would vanish exactly, since the mass-squared difference $\D m_{\tilde \n_R^s}^2 = m_{\tilde \n_R^s}^2 - m_{\n_R^s}^2 = 0$.

 However, SUSY breaking is mediated to our additional sector via the second term in~\eqref{W}. From $\n_L$ and $\tilde h$ in the loop we obtain 
\be 
\label{Dmnusr}
|\D m_{\tilde \n_R^s}|  \sim y_L \L_\text{susy}/(\sqrt{8} \pi) ,
\ee
where $\L_\text{susy}$ denotes the SUSY scale of \order{\text{TeV}}. This is potentially much larger than its supersymmetric mass $\m_s$.
The leading quantum correction to $m_{\phi_x^\pm}$ follows by insertion from the above as
\be
\label{dmphi}
|\d m_{\phi_x^\pm} | \sim \frac{y_\pm y_L}{8 \pi^2} \L_\text{susy}\, .
\ee
Altogether, the requirement of radiative stability yields a condition on the Yukawa couplings
\be
\label{ycondrad}
y_L  < \frac{8 \pi^2}{\sqrt{2}} \frac{q_x g_x}{y_\pm} \frac{m_X}{\L_\text{susy}} \, .
\ee
We will see that this upper bound on the Yukawa couplings could be in tension with the possibility of an effective neutrino self-interaction demanding large Yukawa couplings~\eqref{gnu}.

\textbf{Sparticle mass spectrum}
With conserved R-parity the  lightest supersymmetric particle (LSP) is stable. One important consequence is that it becomes a cosmic relic.
To identify its nature, we consider the mass spectrum of potentially light sparticles:

The active sneutrinos receive their standard masses. Corrections from the second term in~\eqref{W} are of the order of~\eqref{Dmnusr} and thus should be subleading.
The other sterile sneutrino's mass, $m_{\tilde \n_L^s} \sim \m_s$, should receive only subleading corrections, for example, from a $\tilde \phi_x^-$-$\n_R^x$ loop.

The (right-handed) xsneutrino obtains the leading contribution to its mass
\be
m_{\tilde \n_R^x} \sim \text{max}[\m_x, \, \frac{y_\pm}{\sqrt{8} \pi} |\D m_{\tilde \n_R^s}| ]
\ee
either from the superpotential or the $\tilde \phi_x^-$-$\n_L^s$ loop and $\tilde \phi_x^+$-$\bar{\n}_R^s$ loop, respectively. 

Since this loop contribution cannot raise the mass of the (left-handed) xsneutrino, its mass
\be
m_{\tilde \n_L^x} \sim \m_x
\ee
will be similar to its superpartner's mass. In anticipation of~\eqref{gnu} small $\m_x$ is desired for a sizeable neutrino coupling. Consequently, we expect the (left-handed) xsneutrino $\tilde \n_L^x$ to be the LSP in our set up.

\begin{figure}
\label{fig:2loop}
\centering
 \includegraphics[width= 0.9 \columnwidth]{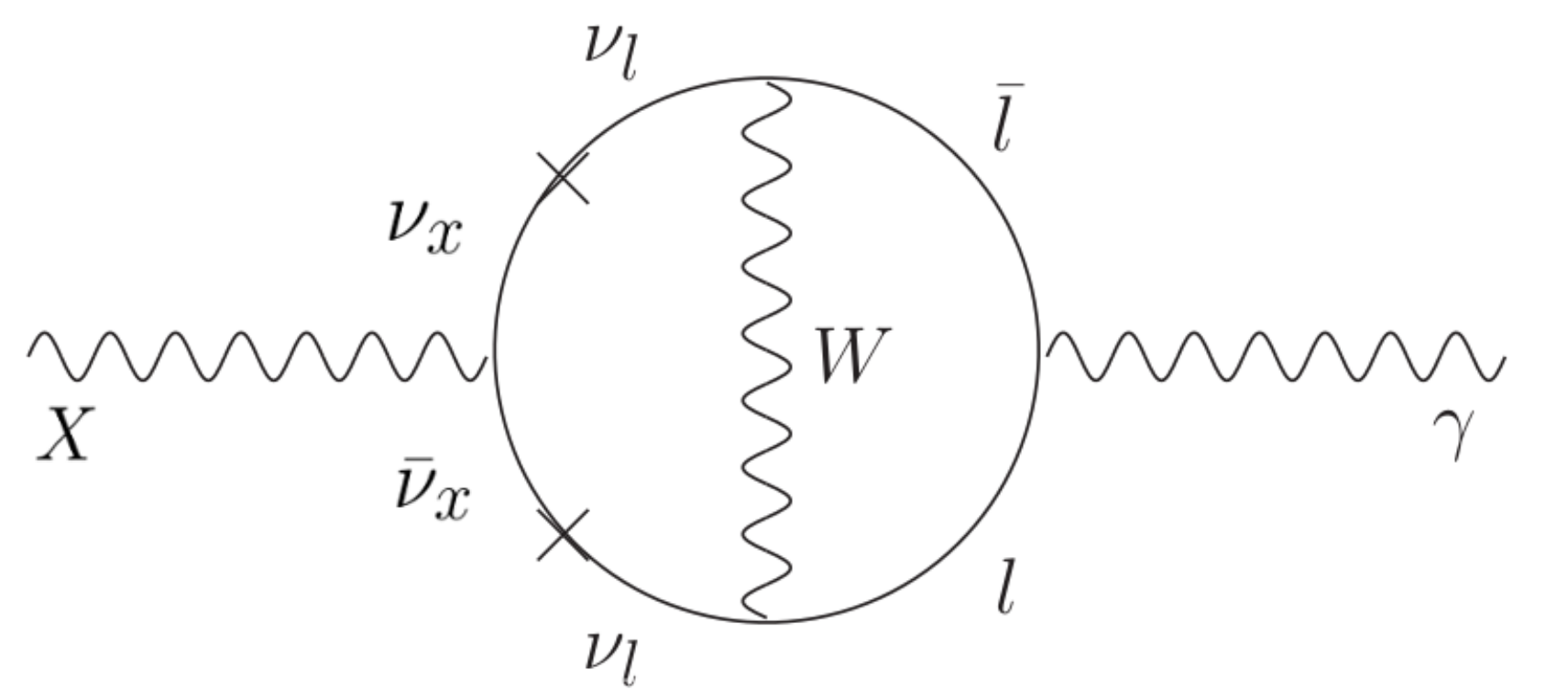}
\caption{Two-loop Feynman diagram generating $X$-$\g$ kinetic mixing.}
\end{figure}

\textbf{Kinetic mixing}
The symmetries allow a kinetic mixing term
\begin{equation}
\label{Lkinmix}
 \mathcal{L}_\text{kin.\@ mix} = -\frac{\e}{2} F_{\m\n}^x F^{\m\n} \,,
\end{equation}
where $F^{\m\n}$ denotes the $U(1)_\text{em}$ field strength.
The mixing parameter $\e$ is constrained to be much smaller than one, see~\cite{Jaeckel:2013ija} for an overview. Since no new symmetry necessarily arises as $\e \rightarrow$, such small values are unnatural from the point of view of t'Hooft. However, it is a renormalisable parameter and we follow the naturalness discussion of kinetic mixing in supersymmetric theories in~\cite{Dienes:1996zr}.
It is a virtue of our theory that no particle is charged under both $U(1)_\text{em}$ and \Uonex.
Therefore, there is no kinetic mixing parameter $\e$ induced unavoidably at the one-loop level, which would represent a serious problem given the observational progress~\cite{Jaeckel:2013ija}.
Actually, kinetic mixing is induced at the two-loop level by the diagram given in Fig.~\ref{fig:2loop}. 
From this diagram we roughly estimate (using $g_x \sim e$)
\begin{align}
\label{epsilon}
\epsilon \sim&
\frac{1}{(16\pi^2)^2} G_F m_X^2 \alpha_\text{em} \sin{\Theta}^2 \nonumber \\
\sim&
10^{-21} \left( \frac{m_X}{1\MeV} \right)^2
\left( \frac{\sin{\Theta}}{10^{-2}} \right)^2 \, ,
\end{align}
where $G_F$ denotes the Fermi constant, $\alpha_\text{em}$ the fine-structure constant and $\Theta$ the neutrino mixing angle.
The order of magnitude of $\e$ in~\eqref{epsilon} lies certainly below all standard bounds. More interestingly, this makes $X$ what has been called a "very dark photon"~\cite{Fradette:2014sza}. 
Kinetic mixing of $X$ with the $Z$ boson is induced (as usual) at one-loop level. However, the corresponding bound due to the observed $Z$-width is very weak and, therefore, surely not constraining.
It is a remarkable property of our theory that $X$-$\g$ mixing arises at a higher loop order than $V$-$Z$ mixing.
While an exploration of this opportunity is beyond the scope of this letter, we would like to note that the very small mixing of $X$ with the photon occurs "accidentally" in our theory and appears unavoidably at the same time.

\section{Effective neutrino charges}
\label{sec:eff}
In the following, we provide the mechanism to equip (massless) active neutrinos with an effective $\Uonex$ charge.

After spontaneous symmetry breaking the additional Yukawa terms in~\eqref{W} give rise to neutrino mass mixing terms.
Defining $\vec{\nu} = (\nu_L \text{, } \nu_L^x \text{, } \nu_L^s \text{, } \nu_R^x \text{, } \nu_R^s)^T $ the new mass terms are
\be
\mathcal{L_\text{$\nu$mass}} = - \bar{\vec{\nu}} \, \mathbf{M} \,  \vec{\nu} \, ,
\ee 
where the mass matrix
\be 
\mathbf{M} = \begin{pmatrix}
  \mathbf{0}_{3\times3} & \mathbf{X}  \\
  \mathbf{X}^T & \mathbf{0}_{2\times2} 
 \end{pmatrix}
\ee
with
\be
\mathbf{X} =
\begin{pmatrix}
  0 & y_L v_\text{ew}/\sqrt{2}  \\
  \m_x & y_+ v_x/\sqrt{2}  \\
  y_- v_x/\sqrt{2} &  \m_s 
 \end{pmatrix}.
\ee
The mass eigenstates are related to the gauge eigenstates by two unitary matrices $\mathbf{U}$ and $\mathbf{V}$ according to
\be
\begin{pmatrix}
\nu_L^1 \\
\nu_L^2 \\
\nu_L^3 
\end{pmatrix} =
\mathbf{U}
\begin{pmatrix}
\nu_L \\
\nu_L^x \\
\nu_L^s 
\end{pmatrix},
\quad
\begin{pmatrix}
\nu_R^2 \\
\nu_R^3 
\end{pmatrix} =
\mathbf{V}
\begin{pmatrix}
\nu_R^x \\
\nu_R^s 
\end{pmatrix}
\ee
that are chosen such that
\be
\mathbf{U}^\dagger \mathbf{X} \mathbf{V} = 
\begin{pmatrix}
  0 & 0  \\
  m_2 & 0  \\
  0 &  m_3 
 \end{pmatrix}
\ee
with positive, real entries. The lightest mass eigenstate is exactly massless or $m_1=0$. This is a direct consequence of the asymmetry of $\mathbf{X}$. Majorana mass terms for $\nu_R^s$ and $\nu_L^s$ in~\eqref{W} could raise $m_1$ to a finite value. Of course, it is possible to add further sterile Majorana neutrinos to implement, for example, a see-saw mechanism generating the observed tiny neutrino masses.
%

For simplicity, we set $y_+ = y_- = y_x$ in the following. This also corresponds to the minimal non-supersymmetric case. It is interesting that we could use the additional freedom of different values for the Yukawa couplings, while doing so would make the discussion less clear only.

In the limit $\m_s \gg y_L v_\text{ew} \gg \m_x \gg y_x v_x$ we find approximately 
\be
\mathbf{U} \simeq
 \begin{pmatrix}
  U_{11} &  \frac{ y_x v_x y_L v_\text{ew}}{2 \m_x \m_s} &  U_{13} \\
  U_{21} & 1 & \frac{y_x v_x}{\sqrt{2} \m_s}  \\
  U_{13} & U_{23} &  1
 \end{pmatrix}, \quad
\mathbf{V} \simeq 
 \begin{pmatrix}
  1 & U_{23}  \\
  U_{23} & 1  
 \end{pmatrix}
\ee
where 
\begin{align}
U_{11}&\simeq   \left(1+ \fb{y_L v_\text{ew}}{\sqrt{2}\m_s}{2}\right)^{-\frac{1}{2}} \simeq 1, \nonumber \\ 
 U_{13}  &\simeq U_{11}[y_L v_\text{ew} \leftrightarrow \sqrt{2} \m_s] \simeq  0.
\end{align}
Here, $\m_s \gg y_L v_\text{ew}$ ensures that the sterile component of the massless state is small.
Since $v_\text{ew} \gg v_x$, we expect $ y_L v_\text{ew} \gg  y_x v_x$.
The relative size of $\m_x$ appears less fixed at this point. 
It is chosen with foresight: A large effective neutrino coupling will require $\m_x$ to be as small as possible. However, we will see in Sec.~\ref{sec:thermal} that successful big bang nucleosynthesis requires a lower bound $\m_x \gtrsim 20 \MeV$, while $20 \MeV > y_x v_x$ typically and probably.
%

In the chosen limit the masses become approximately
\be
m_2 \simeq \m_x  \text{, } \quad m_3 \simeq \m_s \, 
\ee
and the effective coupling $g_\n$ of the massless neutrino is given by 
\be
\label{gnu}
g_\n = g_x q_x U_{12} \simeq g_x q_x \frac{y_xv_x y_L v_\text{ew}}{2 \m_x \m_s}. 
\ee
{\bf Cosmogenic neutrino cascades}
Regarding the observational status with only a few neutrinos detected above $100 \TeV$, it might be too soon to explore any parameter space in detail. Instead, we demonstrate how the neutrino cascade solution proposed in~\cite{Ng:2014pca,Ioka:2014kca} can be implemented in our theory providing an example as a proof of principle.

To affect the neutrino propagation, it is reasonable to require an optical depth $\tau \sim 1$ or, in other words, a mean free path smaller than the distance from the source to the detector. For cosmogenic neutrinos this travel distance is of the order of the Hubble length $c/H_0\simeq 4 \text{ Gpc}$. The opacity needs to be tuned to peak around EeV, such that cosmogenic neutrinos become attenuated in a number of scatterings to PeV energy by up-scattering cosmic background neutrinos. It has been shown that for $m_X \sim 100 \MeV$ and large $g_\n \sim 0.3$ the PeV neutrino flux is comparable to the cosmogenic neutrino flux~\cite{Ng:2014pca,Ioka:2014kca}. This would provide an explanation for the spectrum and the aforementioned coincidence problem. 

Questioning the applicability of our setup, it is important to see that: i) the non-relativistic background neutrinos just scatter as mixed mass eigenstates following our discussion above and ii) ultra-high energy cosmic rays do emit active neutrinos, so that these neutrinos need to oscillate before they can scatter via $\Uonex$ interactions. The decisive oscillation length, $l_\text{osc}\sim E/(\D m^2)\sim $ meter, is indeed macroscopic. However, it is much shorter than for the standard neutrino oscillations, which have oscillation lengths that are already much shorter than the cosmic travel distance. So even for thousands of scatterings between emission and detection the neutrino "flavours" should equilibrate  on average before each scattering. The average interaction strength will thus be given indeed by~\eqref{gnu}, which just takes into account the mixing angles.

To be conservative, proposals were made assuming only one charged neutrino state, so we do, too.  It is known, of course, that at least two neutrinos need to have a finite mass. Since we expect effects from standard neutrino oscillations to average out, we do not complicate the discussion by including them.
Altogether, even though additional "flavour" oscillations are involved, our setup appears applicable, so that we can implement the phenomenologically required interaction strength by consideration of the mixing matrix.

It follows straightforwardly from~\eqref{gnu} that the idealizing large hierarchy necessary to perform the analytic approximation cannot lead to sizeable neutrino interactions.
Therefore, we are going to demonstrate in an explicit example that the cascade mechanism can be implemented with only generic and simple values of dimensionless parameters.

Evaluating numerically the exact mixing matrices for $m_X= 100 \MeV$, $\m_s = 5 \GeV$, $\m_x = 100 \MeV$ and $q_x=1$, $g_x=1$, $y_x= y_+= y_-=\sqrt{2}$, $y_L=0.01\times \sqrt{2}$ we find
\be
\mathbf{U} \simeq
 \begin{pmatrix}
 -0.82 &  -0.37 &  0.44 \\
  -0.41 & 0.91 & 0.018  \\
  0.41 & 0.17 &  0.90
 \end{pmatrix}, \quad
\mathbf{V} \simeq 
 \begin{pmatrix}
  -1 & 0.016  \\
  0.016 & 1  
 \end{pmatrix}
\ee
and $g_\n =0.37$. The corresponding particle masses are $m_3 = 5.6 \GeV$ and $m_2 = 108 \MeV$.

We would like to point out: i) it is by far non-trivial that there is an implementation, ii) even though there are requirements, which are in fundamental tension, simple and generic values for dimensionless parameters suffice, and iii) at the same time, particle masses and parameter values are everything but arbitrary. 

\section{Thermal history}
\label{sec:thermal}
At high enough temperatures $\gg$ TeV the whole particle content of the theory is in thermal equilibrium, since the Yukawa terms in~\eqref{W} provide good thermal contact among the different sectors.
As the Universe cools, it passes the TeV and electroweak scale. The MSSM particles annihilate and decay without any relic density. Electroweak symmetry breaks as usual. 
Around the new $\Uonex$ scale of \order{100 \MeV}, the new symmetry breaks spontaneously.

Later the superpartner of the lighter massive neutrino $\tilde \nu_2$ with mass $m_2$, which is the LSP as we can see in Sec.~\ref{sec:theory} together with the requirement of an effective cascade mechanism, begins to freeze-out.
We saw that its composition will be dominated (like 90\%) by the xsneutrino component, which is also the most strongly interacting component.

In general, its relic density $\O h^2$ will be given by $\O h^2 \simeq 1.07 \times 10^9 x_\text{fo}/(\s_0 \mplanck)$, where $\mplanck$ denotes the Planck mass, $x_\text{fo}= \ln[0.038 (4/g_\ast^{1/2}) \mplanck m_2 \s_0]$, where we used that the LSP has four internal degrees of freedom, and $\s_0$ is a (velocity $v$ independent) approximation to the weighted annihilation cross-section $\langle \s_\text{A} |v|\rangle$~\cite{Kolb:1990vq}. For simplicity, we fix the number of relativistic degrees of freedom $g_\ast$ to its value before $e^+$-$e^-$-annihilation, $g_\ast =43/4$.
%

For sizeable $y_x$ the dominant annihilation process of the xsneutrino in the non-relativistic regime is s-channel annihilation via the xHiggs $\phi_x^\pm$ into a pair of massless neutrinos, $\tilde \nu_x \bar{\tilde \nu}_x \rightarrow (\phi_x^\pm)^\ast \rightarrow \nu_1 \bar{\nu}_1$, with
\be
\s_0 \sim y_x^2 (g_x q_x)^4 \frac{v_x^2}{m_{\phi_x^\pm}^4} = \frac{y_x^2 g_x^2 q_x^2}{m_x^2},
\ee
where we used $m_{\phi_x^\pm} = m_X = q_x g_x v_x$. For the explicit example above, this implies a $\O_{\tilde \nu_2} h^2 $ that is roughly nine orders of magnitude smaller than the DM density $\O_\text{dm} h^2\simeq 0.11$.

The lightest of the additional particles will be the lighter massive neutrino $\nu_2$ that annihilates and decays, as all additional particles with the exception of the LSP, ultimately into massless neutrinos $\nu_1$. These decays occur cosmologically fast via the $\Uonex$ interaction avoiding dangerous out-of-equilibrium decays involving small parameters, c.p.~\cite{Hasenkamp:2012ii}.
Lastly, the active-like neutrinos decouple as usual around $T\sim \MeV$.

Altogether, the implied "standard" thermal history of our theory is not in conflict with any cosmological observable, while it does not provide the DM density.
In the following, we would like to comment on two imposing opportunities that arise from two different alterations to the just outlined thermal history.

{\bf Dark Matter}
While an exploration of the DM opportunities in our theory is beyond the scope of this letter, we would like to note the following:

1) To obtain $\O_{\tilde \nu_2} =\O_\text{dm}$ with a mass between $20 \lesssim m_2/\text{MeV} \lesssim 100 \MeV$  a small $y_x q_x g_x \lesssim 10^{-4.6}$ were required, which decreases  the effective neutrino coupling~\eqref{gnu}. We note in passing that the annihilation via X appears  relatively suppressed, because $m_2^2 < v_x^2$.

2) T-channel annihilation processes via the xsparticles, $\tilde \nu_x \bar{\tilde \nu}_x \rightarrow \tilde X^\ast/\tilde h_x^\ast \rightarrow \nu_1 \bar{\nu}_1$  are dimensionwise of the same order of magnitude, since SUSY is only slightly broken in the xsector.  As these processes do not depend directly on $y_x$, however still as $g_\nu^2$ does, they could dominate for smaller $y_x$. In any case, sizeable neutrino interactions deplete the LSP density.

3) The xcharged LSPs freeze out symmetrically. However, they do not form bound states, because their binding energy $E_b = \a_x^2 m_2/2$ with $\a_x= g_x^2 q_x^2/(4\pi)$ is smaller than the X boson mass and at corresponding times their kinetic energy is even smaller. 

4) Such strongly self-interacting dark matter can be detected directly in dark matter detectors. For DM masses above roughly $10 \GeV$ and a $m_X \sim 50 \MeV$, photon and Z kinetic mixing as small as $\e \sim 10^{-10}$ are already excluded by non-detections~\cite{Kaplinghat:2013yxa}.
Having pointed out that, our LSP's mass is far below the detection threshold. At the same time severe constraints on light DM annihilations from the CMB~\cite{Madhavacheril:2013cna} and diffuse gamma ray emission~\cite{Tavakoli:2013zva} do not apply, because the LSP annihilates into invisible neutrino pairs via its $\Uonex$ interaction.

5) Of course, untouched by these considerations is the possibility to enlarge the particle content to obtain a (self-interacting) CDM relic.
 
It is alluring to mention that the idea of a self-interacting sector containing neutrinos and CDM~\cite{Aarssen:2012fx,Dasgupta:2013zpn,Bringmann:2013vra} attained quite some attention recently and appears --even in the case of eV-sterile neutrinos-- not excluded by cosmological data~\cite{Chu:2015ipa}. While beyond of the scope of this letter, this is an additional set of models our theory should be able to incorporate.

{\bf Hot Dark Matter}
Current determinations of the effective number of relativistic degrees of freedom  during and after the epoch of big bang nucleosynthesis (BBN) are consistent with the standard expectation as well as with unobservable hot dark matter (HDM) densities~\cite{Ade:2015xua}. 
To avoid any potential conflict with these observations, the additional supermultiplets may not contribute to the radiation energy density during BBN or the HDM energy density at late time, while $\Uonex$ charged particles potentially stay in thermal equilibrium until late times.

Maintaining standard expectations with additional particles in thermal equilibrium requires their masses to exceed roughly $20 \MeV$~\cite{Boehm:2012gr}.
Straightforwardly, a smaller  $\m_x$ and thus a lighter $\n_2 \sim \n_x$ to increase the effective neutrino coupling~\eqref{gnu} seems excluded.
Interestingly, this  represents an important lower bound on the masses of the additional particles, while the parameter space preferred from the neutrino cascade mechanism with larger $m_X$ as in our explicit example appears perfectly accessible.

The other way around, if with increased precision there should arise evidence for a tiny HDM admixture (similar to~\cite{Hamann:2013iba}), within our setup this could be due to a small increase in the number density of (active-like) neutrinos due to the existence of $\nu_2$ with $m_2\sim 20 \MeV$.

\section{Summary and conclusions} 
Motivated by the IceCube observations, we presented a minimal, UV-complete theory and prove that the cascade mechanism can be implemented successfully with simple, generic values for all dimensionless parameters.
SUSY reduces the number of possible interaction terms  favourably and fixes the xHiggs mass at a preferred value.
Gauge invariance forbids the Higgs portal.
Since no Standard Model eigenstate is charged under the new interaction, kinetic mixing is induced at two-loop order only. So our theory contains a "very dark photon" scenario.
Interestingly enough, our setup can evade any conflict with the observed thermal history of the Universe. In particular, the observed number of relativistic degrees of freedom may take its standard value and the relic density of the lightest supersymmetric particle is generically much smaller than the dark matter density.

The key notion for effectively charged neutrinos is an asymmetric mass matrix leaving them exactly massless. Only one generation needs to obtain charge via the presented mechanism. Known notions for providing them with finite masses, especially, the see-saw mechanism can still be implemented. Apparently this can be done without any conflict.
Another key notion of this work is that SUSY breaking is mediated to the additional particle sector via loops with the only exception of xHiggs terms.
The stability of the small VEVs sets an upper bound on the Yukawa couplings. In fundamental tension, the size of the (favourably large) effective neutrino coupling depends linearly on the same couplings. 
Altogether, it is by far non-trivial that there is an implementation of the cascade mechanism at all. Astonishingly, simple and generic values for dimensionless parameters suffice, while, at the same time, particle masses and parameter values are everything but arbitrary.

There are imposing opportunities arising: One is to find a cold dark matter candidate within the minimal particle spectrum. The addition of a DM particle implements a self-interacting sector of DM and neutrinos with possibly favourable consequences for the formation of structure in the Universe.
The setup can provide a natural explanation for tiny HDM admixtures formed by active-like neutrinos. Last but not least, it is a fascinating opportunity to equip a eV-sterile neutrino solving the oscillation anomalies with additional interactions to reconcile it with cosmological data.

Finally, we would like to remind that the IceCube experiment is running and its data will likely determine the presence of a cut-off in the near future.

\subsection*{Acknowledgements}
\noindent
I would like to thank Steen Hannestad, Jan Hamann, Neal Weiner, Torsten Bringmann and J\"{o}rn Kersten for valuable discussions.
Furthermore, I would like to thank Florian Staub for help with the implementation of the theory in standard computer codes exploiting his tool SARAH~\cite{Staub:2013tta}.
I acknowledge support from the German Academy of Science through the Leopoldina Fellowship Programme 
grant LPDS 2012-14.

\phantomsection 
\addcontentsline{toc}{chapter}{References}
%

\end{document}